\documentstyle[12pt]{article}

\title{ Poincare  invariance in temporal gauge canonical quantization and
\(\theta\)-vacua }

\author{
 Hisashi Kikuchi
\vspace*{0.6em}\\
{\normalsize \sl Department of Physics}\\
{\normalsize \sl University of California, Riverside}\\
{\normalsize \sl Riverside, CA 92521}
}

\date{February, 1993. UCRHEP-T104}

\begin{document}

\maketitle

\vfil

\begin{abstract}
The Poincare invariance in the  temporal gauge canonical
quantization of QCD is shown manifestly
by verifying the energy-momentum-vector and
angular-momentum-tensor satisfy the \newline
Poincare algebra in the  physical
Hilbert space.
Two different values of \(\theta\) for the $\theta$-term in QCD lagrangian
lead  to different representations of the Poincare group,
which are, however, connected by an unitary transformation.
Thus the parameter \(\theta\) becomes physically irrelevant unless
we can further restrict the physical Hilbert space.
\end{abstract}

\pagebreak

\setlength{\baselineskip}{18pt}

\newcommand{\vevOf}[1]{ \langle 0|\, #1\,| 0\rangle }
\newcommand{\cmtOf}[2]{\left[\, #1,\,#2 \,\right]}
\newcommand{\ket}[1]{| #1 \rangle}
\newcommand{\bra}[1]{\langle #1 |}

\newcommand{\delxy}{\delta( \vec x - \vec y\, )}
\newcommand{\delij}{\delta_{ij}}
\newcommand{\delab}{\delta^{ab}}
\newcommand{\eijk}{\epsilon_{ijk}}
\newcommand{\fabc}{f^{abc}}

\newcommand{\pad}[1]{\partial_#1}
\newcommand{\D}[1]{{\cal D}_#1}
\newcommand{\be}{\begin{equation}}
\newcommand{\ee}{\end{equation}}
\newcommand{\bea}{\begin{eqnarray}}
\newcommand{\eea}{\end{eqnarray}}

\newcommand{\G}[2]{G^#1_{#2}}
\newcommand{\Gdot}[2]{\dot G^#1_{#2}}
\newcommand{\E}[2]{E^#1_#2}
\newcommand{\Edot}[2]{\dot E^#1_#2}
\newcommand{\B}[2]{B^#1_#2}
\newcommand{\Bdot}[2]{\dot B^#1_#2}
\newcommand{\EMT}[2]{\theta^{#1#2}}
\newcommand{\EMTtilde}[2]{\tilde \theta^{#1#2}}
\newcommand{\CM}[2]{\pi^#1_#2}
\newcommand{\ofvec}[1]{(\vec #1\,)}
\newcommand{\GL}[1]{ {\cal G}^#1}
\newcommand{\Gtilde}[1]{\tilde {\cal G}^#1}
\newcommand{\Ptilde}[1]{\tilde P ^#1}
\newcommand{\Mtilde}[2]{\tilde M ^{#1#2}}

\section{Introduction}

The speculation that QCD would have multiple vacua is based on a semiclassical
consideration in terms of Euclidean functional integral.
The instanton solutions are interpreted  as the quantum tunneling between
topologically distinct n-vacua and the true vacuum is expected to be
an appropriate linear combination of them
\cite{Belavin.et.al:1975,'t Hooft:1976,%
Jackiw.et.al:1976,Callan.et.al:1976,Jackiw:1977,Callan.et.al:1978,%
Jackiw:1980}
\be \ket \theta \sim \sum_n e^{i n \theta} \ket n. \ee
As  a result  we obtain a  set of vacua parametrized by a real parameter
\(\theta\). The world of hadrons is regarded as being constructed
on one of them.

The temporal gauge (\(\G a0 = 0\)) have been exclusively used when this
semiclassical speculation is embodied in terms of the canonical quantization.
This gauge fixing condition does not suffer from Gribov's ambiguity
and is the most convenient one  for incorporating properly  large
field fluctuations  like instantons
\cite{Gribov+Singer:1978,Friedman.et.al:1983}.
In this gauge, however, the Lorentz invariance or, more generally, the
Poincare invariance of the theory is not manifest.
For the multiple vacua to be acceptable theoretically,
 there must be a representation of Poincare group
in each \(\theta\)-sector, the Hilbert space constructed based on each
$\theta$-vacuum.
In this paper, we will verify the Poincare invariance of
 the temporal gauge quantization and discuss about the multiple vacua
speculation from the canonical quantization view point.

The standard way to take into account the effect
of the \(\theta\)-vacuum is to add the Pontryagin density
to the lagrangian \cite{Callan.et.al:1976}.
We will canonically quantize the
lagrangian with the \(\theta\)-term and
show the Poincare invariance within the physical
space, the subspace of the Hilbert space specified by the requirement
of gauge invariance.\footnote{%
A similar investigation in the axial gauge have been done in
Ref.~\cite{Bars.et.al:1978}.}
The important observation that will be shown in this paper is
that different values of \(\theta\) in the lagrangian do not
automatically lead us to different \(\theta\)-sectors.
Instead, they  result in the
different representations of the Poincare group in the physical space.
These representations  are connected by certain unitary transformations.
Thus,  all different values of \(\theta\) make the same
physical predictions.
In order to confirm the semiclassical speculation on the multiple vacua and
examine the \(\theta\)-dependence of QCD,
it is important to explicitly write ``large'' gauge transformations
in terms of field operators.

The Poincare invariance verified  in this paper also gives a firm
theoretical basis to a recently
formulated Lorentz invariant sum rule for  the $\theta$-dependence,
the $\theta$-dependence of the vacuum energy in terms of
a sum over matrix elements of a field operator \cite{Kikuchi.et.al:1992}.

We will ignore quark contribution and consider  pure
gluonic QCD.
This  paper is organized as  follows.
The section 2 contains a review of the quantization method and description
of various notations.
Strictly speaking, the gauge we will use  is not the
usual temporal gauge. We will not set the temporal components \(\G a0\) of
gluon fields to be zero but fix them as arbitrary given c-number functions.
We use the term ``temporal gauge'' in this more general sense.
The advantage of keeping \(\G a0\) nonzero is that equations
are written in more Poincare covariant fashion than setting  \(\G a0 = 0\).
In section 3, we explicitly  verify that the energy-momentum-vector and
angular-momentum-tensor obey the Poincare algebra when restricted in
the physical space.
At this point we assume the existence of at least one vacuum state that
is invariant under spatial translation and rotation.
Section 4 is devoted for discussion.
The effect of  \(\G a0\) and  \(\theta\) on physical observables
will be discussed.

\section{Yang-Mills equations}
The familiar problem in quantizing gauge theory in the canonical hamiltonian
formalism is the absence of momenta conjugate to the
temporal components \(\G a0 \). Then we cannot determine the time evolution
of \(\G a0\) by the Heisenberg equations.
The way we adopt in order to circumvent this problem is simple; we
take \( \G a0 \) as given c-number functions  of  space-time coordinates $x$
and check the independence of physical
observables from \(\G a0 \) at the end of the quantization procedure.
Since we fix \(\G a0\), we no longer have the freedom of the time
dependent gauge transformation.
The canonical equations of motion for the spatial components $\G ai$ and
their conjugate momenta $\CM ai$ form now
a closed set in the sense that
the first order differential equations  completely determine
$\G ai(x)$ and $\CM ai(x)$ for all \(x^0\)  from any given
initial configuration  at $x^0 = 0$.

The QCD lagrangian with the topological \(\theta\)-term reads%
\footnote{Notations:
The Minkowski indices are denoted by Greek letters.
Their lowering and raising is done by the metric \(g^{\mu\nu}=g_{\mu\nu}
= \mbox{diag}( 1, -1, -1, -1 )\).
\( \epsilon_{\mu\nu\lambda\sigma} \) is the four dimensional
completely antisymmetric tensor; \( \epsilon_{0123} = - \epsilon^{0123} = 1\).
Roman letters  $i, j, k ...$ are for spatial
indices and run from 1 to 3.
Three dimensional \(\eijk\) is defined by
\(\epsilon_{123} = 1\).
Roman letters \( a, b, c ...\) denote  the indices of the SU(3) adjoint
representation; they run from 1 to 8.
\(\fabc \) are the SU(3) structure constants and
completely antisymmetric with respect to \(a\),\(b\), and \(c\).
Repeated indices of every type are summed over.}
\begin{equation}
{\cal L}  = - {1\over 4} \G a{\mu\nu} G^{a\mu\nu} + \theta {g^2 \over 32\pi^2}
\G a{\mu\nu} \tilde G^{a\mu\nu}, \label{eq:LQCD}
\end{equation}
where
\begin{equation}
\G a{\mu\nu} = \pad \mu \G a{\nu} - \pad \nu \G a{\mu} + g \fabc
\G b\mu \G c\nu, \qquad \tilde G^{a\mu\nu} = {1\over 2}
\epsilon^{\mu\nu\lambda\sigma} \G a{\lambda\sigma},
\end{equation}
and \(g\) is the coupling constant.
Following the usual procedure to get the canonical form of the theory,
we calculate conjugate momenta
\begin{equation}
\CM ai \equiv { \partial {\cal L} \over \partial \Gdot ai }
= \Gdot ai - \D i\G a0
- \bar\theta \B ai,
\end{equation}
where \( \B ai \) stand for chromo-magnetic fields,
\begin{equation}
\B ai = {1\over 2} \eijk \G a{jk},\label{eq:B}
\end{equation}
\(\D \mu\) the covariant derivative for adjoint representation, for example,
\begin{equation}
\D \mu A^a \equiv \pad \mu A^a + g \fabc \G b\mu A^c,
\end{equation}
and \(\bar\theta \equiv ( g^2 / 8\pi^2 ) \theta \).
The hamiltonian is then given by
\begin{equation}
\tilde {H} \equiv  \int d\vec x \left( \CM ai\Gdot ai - {\cal L} \right) =
\int d\vec x \left[ {1\over 2} \left( \CM ai + \bar\theta \B ai \right)^2 +
{1\over 2} \B ai{}^2 + \CM ai ( \D i \G a0 ) \right], \label{eq:tilH}
\end{equation}
where the tilde is used to distinguish \( \tilde H \) from its gauge invariant
counterpart \(H\); see equations below.

The  equations of motion for the canonical fields are the Heisenberg equations
\begin{eqnarray}
\Gdot ai ( x) & = & i \cmtOf{ \tilde H }{ \G ai( x) }
\label{eq:Heq1}\\
\dot \CM ai ( x) & = & i \cmtOf{ \tilde H }{ \CM ai ( x ) }
\label{eq:Heq2}
\end{eqnarray}
in terms of the equal time commutation relations.
The hamiltonian has an explicit \(x^0\)-dependence through $\G a0$.
$\tilde H$ at the same $x^0$ is used for  determining the
time evolution of $\G ai(x)$ and $\CM ai(x)$ in (\ref{eq:Heq1})
and (\ref{eq:Heq2}).
[Throughout this paper, our commutator is taken only at  equal time.
Thus  two  operators in commutators are understood to have the same
time argument
 if they have $x^0$-dependence.]
We  give initial operator configurations  for the canonical fields
at \( x^0 =0\)
such that they obey the  commutators
\begin{eqnarray}
\cmtOf {\CM ai \ofvec x} {\G bj \ofvec y}
& = & -i \delij\delab\delxy \label{eq:CC1} \\
\cmtOf {\CM ai \ofvec x} {\CM bj \ofvec y}
& = & \cmtOf{\G ai \ofvec x} {\G bj \ofvec y} = 0. \label{eq:CC2}
\end{eqnarray}
The equations (\ref{eq:Heq1}) and (\ref{eq:Heq2})
determine the canonical fields for all  $x$
and the solutions  \( \G ai(x) \) and \(\CM ai (x)\)  obey
(\ref{eq:CC1}) and (\ref{eq:CC2}) for  any \(x^0\).

It is convenient to define the  operator fields \(\E ai\),
chromo-electric field, by
\begin{equation}
\E ai(x)  \equiv \CM ai(x) + \bar\theta \B ai(x) \label{eq:E}
\end{equation}
and to write commutators and Heisenberg equations in terms of
$\G ai$, $\E ai$, and $\B ai$.
Using (\ref{eq:CC1})--(\ref{eq:E}) we obtain
\begin{eqnarray}
\cmtOf{\G ai \ofvec x} {\G bj \ofvec y} & = &
\cmtOf {\B ai \ofvec x}{ \G bj \ofvec y}  =
\cmtOf{ \B ai \ofvec x}{\B bj \ofvec y} = 0, \label{eq:ETC1}\\
\cmtOf{ \E ai \ofvec x}{ \G bj \ofvec y } & = & -i \delij\delab\delxy, \\
\cmtOf{ \E ai \ofvec x}{ \B bj \ofvec y } & = &
i \eijk\left\{ \delab\pad k\delxy - g \fabc \G  c k \ofvec y \delxy \right\} \\
\cmtOf{\E ai \ofvec x}{\E bj \ofvec y} & = & 0,   \label{eq:ETC2}
\end{eqnarray}
while the  hamiltonian  becomes
\begin{equation}
\tilde H = \int d\vec x \left\{
{1\over 2} \E ai {}^2 + {1\over 2} \B ai {}^2 - ( \D i \E ai )\G a0
\right\}.  \label{eq:tH}
\end{equation}
Here we have used the equations
\be \D i \CM  ai = \D i \E ai, \label{eq:ID}\ee
which are the consequences of Bianchi identities
\be \epsilon^{\mu\nu\lambda\sigma} \D\nu \G a {\lambda\sigma} = 0
\label{eq:Bianchi} \ee
for \( \mu = 0 \).
We have also neglected a surface integral at spatial infinity in
Eq.~(\ref{eq:tH}).
In this paper we neglect similar surface integrals under the  assumption
that they cannot change the dynamics.
The Heisenberg equations now read
\bea \Gdot ai &  = & i \cmtOf {\tilde H} { \G ai } = \E ai + \D i \G a0
\label{eq:EforG}\\
\Edot ai & = & i\cmtOf{\tilde H}{\E ai } = -\eijk\D j\B ak
- g\fabc \G b0 \E ci,  \label{eq:EforE}
\eea
and
\be
\Bdot ai = i \cmtOf {\tilde H} {\B ai} = \eijk \D j \E ak - g \fabc
\G b0 \B ci.  \label{eq:EforB}
\ee
Eqs.~(\ref{eq:EforG}) clearly mean operator fields \( \E ai \) defined by
(\ref{eq:E}) can be identified as electric fields \( \G a{0i}  \).
Eqs.~(\ref{eq:EforE})  are  the operator version of
the Yang-Mills equations
\be \D \mu G^{a\mu\nu} = 0 \label{eq:YM} \ee
for \(\nu = i\).
Eqs.~(\ref{eq:EforB}), which  are  consistent with
(\ref{eq:B}) and (\ref{eq:EforG}), are the spatial components of
Bianchi identities (\ref{eq:Bianchi}).

The time components of the Yang-Mills equations,
\be \GL a \equiv \D i\E ai = 0, \label{eq:GL} \ee
cannot be satisfied as an operator equation.
Instead, it should be interpreted as a constraint on  the physical states
$\ket{\mbox{ }} $  \cite{Jackiw:1980,Weyl}. They are defined by
\be  \GL a \ofvec x  \ket{\mbox{ }} = 0 \label{eq:phys} \ee
for all  $a$ and \( \vec x \) with \(\GL a\) at \(x^0 = 0\).
Although the time derivative of $\GL a$ is not zero,
\be \dot \GL a = i \cmtOf{\tilde H} {\GL a} = - g\fabc \G b0 \GL c,  \ee
its matrix elements in the physical space  are zero.
This means (\ref{eq:phys}) holds for arbitrary \(x^0\).

\newcommand{\Ophys}{{\cal O}_{\mbox{\scriptsize ph}}}

For the verification of the Poincare  invariance, we need one more definition.
We refer to  the hermitian operators  that map
the physical states into the physical ones as physical.
A physical operator \(\Ophys\) must satisfy
\be \cmtOf{ \GL a ( x)}{\Ophys} \ket{\mbox{ }} = 0 \ee
for arbitrary $a$, \( x\) and physical \(\ket{\mbox{ }}\).
This definition means that matrix representation of \(\Ophys\) are
block diagonal with respect to the physical space and the unphysical space
(the orthogonal complement of the physical space).
Let us refer to sub-matrix of \(\Ophys\) in the physical space
as the physical component.

The operators \(\GL a\) generate  local gauge transformations:
they satisfy
\begin{eqnarray}
\cmtOf{\GL b\ofvec x}{\G ai\ofvec y} &=& -i
\left\{g\fabc \G ci \ofvec y \delxy + \delab\pad i \delxy \right\},\\
\cmtOf{\GL b\ofvec x}{\E ai\ofvec y} & = &
-i  g\fabc \E ci \ofvec y \delxy, \label{eq:GonE}\\
\cmtOf{\GL b\ofvec x}{\B ai\ofvec y} & = &
-i  g\fabc \B ci \ofvec y \delxy,
\end{eqnarray}
and local algebra of gauge group
\be
\cmtOf{ \GL a \ofvec x}{\GL b \ofvec y} =
ig\fabc\GL c\ofvec y \delxy.
\ee
Thus the gauge invariant operators, the operators that commute with
\(\GL a\),  are physical.

\section{The Poincare algebra}
Now we will explicitly show the Poincare  invariance of our quantization
scheme, i.e. we will show  the existence of Poincare group representation
in the physical space.
We start with the classical expression of the energy momentum tensor
\be
\EMT \mu\nu = - G^{a \mu\lambda} G^{a\nu}{}_\lambda + {1\over 4}
g^{\mu\nu} G^a_{\lambda\sigma} G^{a\lambda\sigma}
\ee
and examine its properties in terms of the  field operators.
The corresponding operator expressions for \(\EMT \mu\nu \) are
\begin{eqnarray}
\EMT 00 (x) & = & {1\over 2} \left\{ \E ai (x) ^2 +
\B ai (x) ^2 \right\} \\
\EMT 0i (x) & = & - \eijk {1\over 2} \left\{ \E aj (x)  \B ak (x)  +
\B ak (x) \E aj(x) \right\} \\
\EMT ij (x) & = & {1\over 2} \delta_{ij} \left\{ \E ai (x) ^2 +
\B ai (x) ^2 \right\}  - \E ai(x) \E aj(x)  -\B ai(x)  \B aj(x),
\end{eqnarray}
where we have symmetrized the expression for \(\EMT 0i \) with respect to the
order of $\E aj$ and $\B ak$ because they do not commute as operator
fields; we will use the same prescription  for handling the order of operators
when this  becomes a problem.
$\EMT 0i$  satisfy $ \EMT 0i = \EMT i0 $ and are hermitian.
We will prove that the unitary transformations
\be \Lambda(\lambda, \omega) \equiv  \exp\left\{ i \lambda_\mu P^\mu +
i\omega_{\mu\nu} M^{ \mu\nu }
\right\}, \ee
defined by
the energy-momentum-vector  \(P^\mu\)
\be P^\mu  \equiv \int d\vec x \, \EMT 0\mu (x)  \label{eq:P} \ee
and the angular-momentum-tensor $M^{\mu\nu}$
\be
M^{\mu\nu} \equiv
\int d\vec x \, \left( x ^\mu \EMT 0\nu (x) - x^ \nu \EMT 0\mu (x) \right)
\label{eq:M}
\ee
with real parameters $\lambda_\mu $ and  $\omega_{\mu\nu} $,
constitute a representation of the  Poincare group in the physical space.

The operators $\EMT \mu\nu$ are gauge invariant and, thus, physical.
They satisfy divergence equations
\bea  \dot \EMT 00 + \pad j \EMT 0j & = & 0  \\
\dot \EMT 0i + \pad j \EMT ji  & = & - \E ai \GL a, \eea
which are the simple consequence of  (\ref{eq:EforE}) and (\ref{eq:EforB}).
Note that the  operators $ \E ai \GL a $ are hermitian and physical:
$\E ai $ and $\GL a$ commute for the same color index $a$ no matter what their
spatial coordinates are (See Eq.~(\ref{eq:GonE})).
And their physical components are zero:
\be  \E ai \GL a \ket{\mbox{ }}  = 0 \quad
\mbox{for any physical } \ket{\mbox{ }}. \ee
Thus, although  some of \(P^\mu\) and $M^{\mu\nu}$
 (which have \(\EMT 0i\) in their expressions) are $x^0$-dependent,
their physical components  are $x^0$-independent.
The unitary transformations  $ \Lambda(\lambda, \omega)$ then
 have an unique operation in the physical space
no matter what time slice in the Minkowski space we use for evaluating the
integrals  (\ref{eq:P}) and (\ref{eq:M}).

The evaluation of commutators of \(\EMT \mu\nu \) with respect to the
operator fields are straightforward. By Eqs.~(\ref{eq:ETC1})--(\ref{eq:ETC2}),
we obtain
\begin{eqnarray}
\cmtOf{\EMT 00 \ofvec x}{\G a i \ofvec y} & = &-i \E ai\ofvec y \delxy
\label{eq:COfEMT1}\\
\cmtOf{\EMT 0k \ofvec x}{\G a i \ofvec y} & = & i \eijk\B a j\ofvec y\delxy
\label{eq:example}\\
\cmtOf{\EMT 00\ofvec x}{\E ai\ofvec y} & = &
i \eijk \left\{ \left(\D j\B ak\ofvec y\right)\delxy
 - \B ak\ofvec y \pad j \delxy \right\} \nonumber\\
\\
\cmtOf{\EMT 0k\ofvec x}{\E ai\ofvec y} & = &
i ( \delta_{km} \delta_{il} - \delta_{ki}\delta_{lm} ) \nonumber\\
&& \times
\left\{\left(\D m\E al\ofvec y\right) \delxy
- \E al\ofvec y\pad m\delxy\right\} \nonumber\\
\\
\cmtOf{\EMT 00\ofvec x}{\B ai\ofvec y} &= & - i\eijk
\left\{ \left(\D j \E ak\ofvec y\right)  \delxy
- \E ak\ofvec y \pad j\delxy \right\} \nonumber\\
\\
\cmtOf{\EMT 0k\ofvec x}{\B ai\ofvec y} & =&
i ( \delta_{kj} \delta_{il} - \delta_{ki}\delta_{lj} ) \nonumber\\
&&\times
\left\{\left( \D j\B al\ofvec y \right) \delxy
- \B al\ofvec y \pad j\delxy \right\}. \label{eq:COfEMT2}
\end{eqnarray}
Using these results, we can  calculate the commutators of \(P^\mu\).
But these are {\em not} the commutators the generators of translations
should obey;
for example, from (\ref{eq:example}) we get
\be
\cmtOf {P^k} {\G ai \ofvec x}  =  - i \partial^k \G ai - i \D i \G ak, \ee
which has the residual second term on the right hand side.

A careful inspection on Eqs.~(\ref{eq:COfEMT1})--(\ref{eq:COfEMT2}),
however, tells us a simple modification on \(\EMT 0\mu  \) gives us a
right form for the generators of  translations.
We define
\be \EMTtilde 0\mu = \EMT 0\mu - \Gtilde \mu \ee
with hermitian $\Gtilde \mu$,
\be \Gtilde \mu (x) \equiv {1\over 2} \left\{ \GL a (x) G^{a\mu}(x) +
G^{a\mu}(x) \GL a (x) \right\}. \label{eq:Gtil} \ee
The commutators of \(\EMTtilde 0\mu \) are then
\bea
\cmtOf{\EMTtilde 00 \ofvec x}{\G a i \ofvec y} & = &-i \left\{ \E ai\ofvec y
+ \D i \G a0 \ofvec y \right\} \delxy + i \G a0 \ofvec y \pad i \delxy
,\nonumber\\
\label{eq:COftEMT1}\\
\cmtOf{\EMTtilde 0k \ofvec x}{\G a i \ofvec y} & = &
i \left( \pad k \G ai \ofvec y \right) \delxy -i \G ak\ofvec y
\pad i \delxy, \\
\cmtOf{\EMTtilde 00\ofvec x}{\E ai\ofvec y} & = &
i \eijk \left\{ \left(\D j\B ak\ofvec y\right)\delxy
 - \B ak\ofvec y \pad j \delxy \right\}
\nonumber\\
&& + ig\fabc \G b0 \ofvec y \E ci \ofvec y \delxy, \\
\cmtOf{\EMTtilde 0k\ofvec x}{\E ai\ofvec y} & = &
i \left( \pad k \E ai \ofvec y \right) \delxy  \nonumber\\
&& - i ( \delta_{km} \delta_{il} - \delta_{ki}\delta_{lm} )
\E al\ofvec y \pad m \delxy,
\\
\cmtOf{\EMTtilde 00\ofvec x}{\B ai\ofvec y} &= & - i\eijk
\left\{ \left(\D j \E ak\ofvec y\right)  \delxy
- \E ak\ofvec y \pad j\delxy \right\}  \nonumber\\
&& +ig \fabc \G b0 \ofvec y \B ci \ofvec y \delxy,
\\
\cmtOf{\EMTtilde 0k\ofvec x}{\B ai\ofvec y} & =&
i\left( \pad k \B ai \ofvec y \right) \delxy \nonumber \\
&& - i ( \delta_{kj} \delta_{il} - \delta_{ki}\delta_{lj} )
\B al\ofvec y \pad j\delxy. \label{eq:COftEMT2}
\eea
Thus the operators $\Ptilde \mu$, defined in similar way to (\ref{eq:P})
with $\EMTtilde 0\mu$, satisfy
\bea
\cmtOf {\Ptilde \mu} {\G ai ( x)} & = & - i \partial^\mu \G ai ( x),
\label{eq:trG}\\
\cmtOf {\Ptilde \mu} {\E ai ( x)} & = & - i \partial^\mu \E ai ( x), \\
\cmtOf {\Ptilde \mu} {\B ai ( x)} & = & - i \partial^\mu \B ai ( x).
\label{eq:trB}
\eea
Since \( \Ptilde \mu \) are not \(x^0\)-independent, these relations does
not necessarily mean \(\Ptilde \mu\) are the generators of finite translations.
But at least they generate infinitesimal translations for all operator fields.
Note that \(\Ptilde 0 = \tilde H \) so that Eqs.~(\ref{eq:trG})--(\ref{eq:trB})
for  \(\mu = 0\) are the Heisenberg equations
(\ref{eq:EforG})--(\ref{eq:EforB}).

It is interesting to notice that \(\Mtilde \mu\nu\),
defined by (\ref{eq:M}) with
$\theta ^{\mu\nu} $ replaced by $\EMTtilde \mu\nu$, also  satisfy the
commutation relations of the generators for infinitesimal Lorentz
transformations.
Indeed, we obtain
\bea
\cmtOf{\Mtilde \mu\nu}{\G ai (x)} = - i \Delta^{[\mu\nu]} \G ai (x), \\
\cmtOf{\Mtilde \mu\nu}{\E ai (x)} = - i \Delta^{[\mu\nu]} \E ai (x),
\label{eq:rE}\\
\cmtOf{\Mtilde \mu\nu}{\B ai (x)} = - i \Delta^{[\mu\nu]} \B ai (x),
\label{eq:rB}
\eea
where we have introduced a compact notation \(\Delta^{[\alpha\beta]}\) for
infinitesimal Lorentz transformations. They stand for
\be \Delta^{[\alpha\beta]} \G a\mu = \chi^{[\alpha\beta]}_{\mu\nu}
G^{a\nu} + \chi^{[\alpha\beta]}_{\lambda\sigma}x^\lambda \partial^\sigma
\G a\mu, \ee
for the vector potential and
\be \Delta^{[\alpha\beta]} \G a{ \mu\nu}  =
\chi^{[\alpha\beta]}_{\mu\lambda} G^{a\lambda}{}_\nu +
\chi^{[\alpha\beta]}_{\nu\lambda} G^a{}_\mu{}^\lambda{} +
\chi^{[\alpha\beta]}_{\lambda\sigma}x^\lambda \partial^\sigma
\G a {\mu\nu}
\ee
for the field strength; \(\chi^{[\alpha\beta]}_{\mu\nu} \)
is the anti-symmetric parameter for the \([\alpha\beta]\)-Lorentz
transformations,
\be \chi^{[\alpha\beta]}_{\mu\nu} = \delta^\alpha_\mu \delta^\beta_\nu -
\delta^\alpha_\nu \delta^\beta_\mu.
\ee
Note further that although $\Gtilde \mu$ are not gauge invariant,
they are physical:
\be \cmtOf{ \GL a ( x)}{\Gtilde k ( y)} =
 i \GL a (y) \pad k\delxy. \ee
Thus  \(\Ptilde \mu \) and  \(\Mtilde \mu\nu \) are also
physical.

Now we will prove  that physical components of \(\Ptilde \mu \) and
\(\Mtilde \mu\nu\) are the same as those of $P^\mu$ and
$M^{\mu\nu}$, respectively.
We assume that there is at least one vacuum state $\ket 0$ in the physical
space and that it is invariant under spatial translations and rotations
generated by \(\Ptilde i\) and \(\Mtilde jk\) evaluated at \(x^0 = 0\).
Note that finite transformations
\be \tilde \Lambda (\lambda, \omega)
= \exp \left\{ i \lambda_i \Ptilde i + i\omega_{jk}
\Mtilde jk \right\} \ee
do not change the time coordinate when operating on the field operators.
Thus the infinitesimal form that \(\Ptilde i\) and \(\Mtilde jk\)
satisfy, (\ref{eq:trG})--(\ref{eq:rB}), are sufficient
to prove that \(\tilde \Lambda \) represent three dimensional
  translation-rotation group.
Under the above assumption  we will prove
\be \Gtilde \mu ( x)  \ket{\mbox{ }}=  0 \label{eq:Gmu} \ee
for arbitrary physical \(\ket{\mbox{ }}\) and $x$.

For  \(\mu = 0\), (\ref{eq:Gmu}) is  obvious:
since $\G a0$ is c-number function, the definitions
(\ref{eq:phys}) for physical states and (\ref{eq:Gtil}) for
$\Gtilde 0$ result in it.
For $\mu = i$, the time derivative of $\Gtilde i$,
\be \dot {\Gtilde i} = - \left( \E ai + \pad i \G a0 \right) \GL a, \ee
has zero physical components:
\be \dot {\Gtilde i} \ket{\mbox{ }} = 0. \ee
Thus we only need to prove (\ref{eq:Gmu}) at $x^0 = 0$.
Let us write $\Gtilde i$ at $x^0 = 0$ as
\be \Gtilde i \ofvec x = c^i - \G ai \ofvec x \GL a \ofvec x, \ee
where
\be c^i \equiv  - { 1\over 2} \cmtOf{\GL a \ofvec x }{\G ai\ofvec x }
= \left. {8\over 2} i \,\pad i \delta\ofvec x \right|_{\vec x = 0}. \ee
Although \( c^i\) has ill-defined expression and needs an appropriate
regularization,
it is at most a c-number.  Thus we can choose the vacuum \(\ket 0\)
in order to determine \( c^i\):
\be \Gtilde i \ofvec x  \ket 0 = c^i \ket 0. \ee
Applying \(\tilde \Lambda\) on both side of this equation and
 using the assumption that \(\ket 0\) is invariant under \(\tilde\Lambda \),
we get \( c^i = R^i{}_j\,c^j \)
for an arbitrary SO(3) matrix $R^i{}_j$; that is, $c^i = 0$.
Thus the physical components of \(\Gtilde \mu (x) \) are  zero and
\(P^\mu\) and \(\Ptilde \mu\) or \(M^{\mu\nu}\) and \(\Mtilde \mu\nu\) have
the same physical components.

Using (\ref{eq:trG})--(\ref{eq:rB}) and \( H \equiv P^0 \), we obtain
\bea
\cmtOf{\tilde H }{ H } & = & \cmtOf{\Ptilde j} H
= \cmtOf {\Ptilde j} {P^k} = 0,
\label{eq:Po1}\\
\cmtOf{\Mtilde jk} H & = & 0,
\nonumber \\
\cmtOf{\Mtilde jk}{P^i}  & = &  i \left( \delta_{ij} P^k
- \delta_{ik} P^j\right),
 \nonumber \\
\cmtOf{\Mtilde 0k} H & = & -i P^k,
\nonumber\\
\cmtOf{\Mtilde 0k} {P^i} & = & -i \delta_{ik} H -i\int d\vec x x^k
\E ai \GL a,
\label{eq:Po2}\\
\cmtOf{\Mtilde jk}{M^{lm}} & =& i\left(
\delta_{jl}M^{km} - \delta_{jm}M^{kl} +
\delta_{km}M^{jl} - \delta_{kl}M^{jm}\right),
\nonumber\\
\cmtOf{\Mtilde 0k} {M^{ij}} & = & i \left( \delta_{jk} M^{0i}  - \delta_{ik}
M^{0j} \right) -i \int d\vec x x^k \left( x^i \E aj \GL a- x^j \E ai \GL a
\right),
\nonumber \\
\cmtOf{\Mtilde 0k} {M^{0i}} & = & i M^{ik}
 -i \int d\vec x x^0 x^k \E ai \GL a. \label{eq:Po3}
\eea
Since the physical operators have no matrix elements between
physical states and unphysical states, the relations
(\ref{eq:Po1})--(\ref{eq:Po3}) are correct for their physical components as
well.
Recall $\E ai \GL a$ has zero physical components.
Thus the physical components  of \( M^{\mu\nu} (\Mtilde \mu\nu) \) and
\(P^\mu (\Ptilde \mu )\) are $x^0$-independent%
\footnote{
Although the physical component of \(M^{0k} \) does not commute
with that  of \(\tilde  H \), the explicit
\(x^0\)-dependence in its definition cancel the \(x^0\)-dependence
from the commutator in the Heisenberg equations.}
and satisfy
the Poincare algebra.
The operations of the unitary transformations $\Lambda(\lambda, \omega)$
in the physical space are unambiguously determined, and they represent
the Poincare group.

\section{Discussions}
Using the fact that physical components of $P^\mu$ commute each other,
we can  obtain basis vectors in physical space as eigen vectors of
them.
Let us refer this basis as physical basis.
The set of physical  basis and operator solutions for the operator fields
$\CM ai (x)$ (or $\E ai (x)$) and $\G ai(x)$ are the all we need to know
to make  physical predictions.
Especially, the matrix elements of physical operators with respect to
the physical basis are related to gauge invariant physical predictions.

We first discuss about the effect of \(\G a0\) on physical observables.
Since \(\G a0\) is an arbitrary function of the space time coordinate \( x\),
it should have introduced violation of the Poincare invariance had  it
coupled to a physical degree of freedom.
Conversely, our manifest  construction of the Poincare generators
implies that  physical predictions are independent from the
specific configuration of \(\G a0\) that we fix when we start
the quantization procedure.
This statement is assured by the following two things.
1) The physical basis is  \(\G a0\)-independent:
they are the  eigen vectors of the \(x^0\)-independent physical component of
\(P^\mu\) and they are determined at \(x^0 = 0\) by
the initial configurations  of \(\CM ai\) and \(\G ai\)
and the parameter \(\theta\).
2) The time evolution of the physical components of physical operators
\(\Ophys\) is \(\G a0\)-independent: \(\G a0\) can only couple through the
 Heisenberg equation
\be \dot \Ophys = i \cmtOf{ \tilde H}{ \Ophys} \ee
where they  always appear in conjunction with  \(\GL a\),
which have zero physical components.
Thus the matrix elements of \(\Ophys\) with respect to the physical
basis are \(\G a0\)-independent.

Now let us turn to the \(\theta\)-dependence.
We notice that all the \(\theta\)-dependence concentrates in the
definition of \( \E ai\), (\ref{eq:E});
although we started with the lagrangian with \(\theta\)-term,
there is no explicit \(\theta\)-dependence in
the commutators (\ref{eq:ETC1})--(\ref{eq:ETC2}),  hamiltonian  (\ref{eq:tH}),
or the Heisenberg equations (\ref{eq:EforG})--(\ref{eq:EforB}).
Thus we will intensively consider the effect of changing \(\theta\) in
(\ref{eq:E}).
Specifically, we take a following picture for  this consideration.
We think the canonical fields \(\CM ai \ofvec x\) (at \(x^0 = 0\)) have
\(\theta\)-independent operations in the Hilbert space
as the operators that act on the state vectors.
The operators \(\E ai\ofvec x \),
therefore, have \(\theta\)-dependent operations.
(See figure 1 where we represent \(\E ai\) at different values of \(\theta\)
by  arrows as they map a
state vector to another one.)
The physical space does not change for all values of \(\theta\): because of
Eq.~(\ref{eq:ID}), all the constraints on the physical space
for different \(\theta\) are identically \(\D i\CM ai\ket{\mbox{ }} = 0 \).
Now our problem is how physical predictions depend on the
\(\theta\)-dependent  initial operator configurations \(\E ai\ofvec x\)
while  the physical space is \(\theta\)-independent and
the equations of motion do not have explicit  \(\theta\)-dependence.

The key for answering this question is a transformation
\be T ({\varphi} ) \equiv e^{i{\varphi } q } \ee
defined by the so-called topological charge
\be q \equiv \left. \int d\vec  x\,  K^0(x)\right|_{x^0 = 0} \ee
and real parameter \(\varphi\).
Here  \( K^0\) is the time component of the current
\be K^\mu = {g^2 \over 32\pi^2 }
\epsilon^{\mu\nu\lambda\sigma}
\left( \G a\nu \G a{\lambda\sigma} - { g\over 3} \fabc \G a\nu\G b\lambda
\G c\sigma \right)
\ee
whose divergence is the Pontryagin density $\pad \mu K^\mu = (g^2 /32\pi^2)
\G a{\mu\nu}\tilde G ^{a\mu\nu} $.
[This divergence equation holds even for the operator fields
by simply setting  $\G a{0i} = \E ai $ and using (\ref{eq:EforG}).]
\(T(\varphi)\) transforms the operator fields (at \(x^0 = 0\)) as
\bea
T( {\varphi} ) \E ai\ofvec x  T( {\varphi} )^{-1}
& = & \E ai \ofvec x + {g^2\over 8\pi^2} {\varphi} \B ai\ofvec x, \\
T( {\varphi} ) \G ai\ofvec x T( {\varphi} )^{-1} & = & \G ai\ofvec x,
\eea
i.e., it connects the initial operator configurations at \(\theta\) with
those at \(\theta + \varphi\).
Let  \({\cal O}_\theta (x) \)  denote collectively the operator solutions,
\(\E ai(x) \) and \(\G ai(x) \), for the Heisenberg equations with
the initial operator configurations  at \(\theta\).
Then define
\be
{\cal O}_{\theta + \varphi }(x) = T (\varphi) {\cal O}_\theta (x) T
(\varphi)^{-1}. \label{eq:rop}
\ee
\({\cal O}_{\theta +\varphi }(x) \) also satisfies  the equations of motion
 (\ref{eq:EforG})--(\ref{eq:EforE}) and has the appropriate initial
configurations for \(\theta + \varphi\).
[Note that time evolution of \( {\cal O}_{\theta + \varphi }\)
is governed by the hamiltonian written in terms of
\({\cal O}_{\theta+\varphi}\) themselves.]
Thus the operator fields at different values of \(\theta\) are related by
(\ref{eq:rop}).

\newcommand{\emptyket}{\ket{\mbox{ }}}

Next, we consider the effect of changing \(\theta\) on the physical basis.
The operators \( P^\mu\) and \(M^{\mu\nu}\) are written in terms of \(\E ai\)
(and \(\B ai\)) and, thus, they also have \(\theta\)-dependent operations.
Especially different values of \(\theta\) lead us to different representations
of the Poincare group and different sets of  physical basis.
Obviously, \(P^\mu\) or \(M^{\mu\nu}\) at \(\theta\) and \(\theta +\varphi\)
are related by
\bea
P^\mu_{\theta +\varphi} = T(\varphi) P^\mu_\theta T(\varphi)^{-1},\\
M^{\mu\nu}_{\theta +\varphi} = T(\varphi) M^{\mu\nu}_\theta T(\varphi)^{-1}.
\eea
Since \( q \) is physical,
\be \cmtOf{ \GL a \ofvec x } {q} =
i {g^2 \over 32\pi^2 } \int d\vec y\, \eijk (\pad j \G ak\ofvec y ) \pad i
\delxy = 0, \ee
the transformation \(T(\varphi)\) is unitary within the physical space.
Thus physical basis at \(\theta +\varphi \),
\(\emptyket_{\theta +\varphi}\), is related to
  one at \(\theta\), \(\emptyket_\theta\),  by
\be \emptyket_{\theta +\varphi} = T(\varphi) \emptyket_\theta. \label{eq:rket}
\ee
Eqs.~(\ref{eq:rop}) and (\ref{eq:rket}) assert that
two theories with different values of \(\theta\) yield the
same physical predictions and have the same physical content.

The possibility of encountering a physical quantity with a nontrivial
$\theta$-dependence can only occur if we can further
restrict the physical space into a smaller subspace where
the transformation $ T(\varphi) $ loses its unitarity.
By the requirement of Poincare invariance, this restricted space must be
large enough to retain the Poincare algebra generated by $P^\mu$ and
$M^{\mu\nu}$.
Usually assumed in the literature \cite{Jackiw.et.al:1976,Callan.et.al:1976,%
Jackiw:1977,Callan.et.al:1978,Jackiw:1980} is the
existence of the ``large'' gauge
transformation $\Omega$ that transforms the operator fields as
\bea
t^a \, \Omega \G ai\ofvec x \Omega^{-1} & = & h\ofvec x ^{-1} t^a h\ofvec x
 \,\G ai\ofvec x + {i\over g} h\ofvec x ^{-1}  \pad i h\ofvec x,
\label{eq:OmegaG}\\
t^a\, \Omega \E ai\ofvec x \Omega^{-1} & = & h\ofvec x^{-1} t^a h\ofvec x
\, \E ai \ofvec x,\label{eq:OmegaE} \eea
where $t^a$ is the 3\(\times\)3 hermitian traceless generators for SU(3) and
$h\ofvec x $ denotes a representative of local gauge transformation which
has unit winding number as a map S$^3\rightarrow$SU(3).
Once we have the explicit $\Omega$, we can further restrict
the physical space by the requirement
\be \Omega \ket{\mbox{ }} = \ket{\mbox{ } }. \label{eq:res}\ee
Since \(\Omega\) commutes with \(P^\mu\) and \(M^{\mu\nu}\),
the Poincare group is representable within the physical space further
restricted by (\ref{eq:res}). But
\(\Omega\) does not commute with \(q\) and, thus,  \(T(\varphi)\) is no longer
a unitary transformation in this restricted physical space.

$\Omega$ cannot be obtained simply by accumulating infinitesimal
gauge transformations generated by \(\GL a\).
Rather,  the reason why
QCD may have a  nontrivial \(\theta \)-dependence is that the large gauge
transformation is disconnected from those generated by \(\GL a\)
\cite{Jackiw:1980,Wu.et.al:1985}.
As far as we know, a satisfactory operator expression for
\(\Omega\) have not been given yet.
It is important to construct \(\Omega\) explicitly to verify the
multi vacua speculation  and further examine the \(\theta\)-dependence of QCD.

\medskip

\begin{center}
\bf Acknowledgement
\end{center}
The author thanks J. Wudka for helpful discussions.
This work is in part supported by the US Department of Energy under
Contract No. DE-AT03-87ER40327.

\end{document}